\documentclass[a4paper,fleqn,usenatbib]{mnras}
\hypersetup{draft}
\usepackage{newtxtext,newtxmath}

\usepackage[T1]{fontenc}
\usepackage{ae,aecompl}
\usepackage{url}

\usepackage{graphicx}	
\usepackage{amsmath}	
\usepackage{amssymb}	

\title[Radiation dose on exoplanets]{Stellar Proton Event-induced surface radiation dose as a constraint on the habitability of terrestrial exoplanets}

\author[D. Atri]{
Dimitra Atri,$^{1,2,3}$\thanks{E-mail: atri@nyu.edu}
\\
$^{1}$Center for Space Science, New York University Abu Dhabi, PO Box 129188, Saadiyat Island, Abu Dhabi, UAE\\
$^{2}$Department of Physics, New York University Abu Dhabi, PO Box 129188, Saadiyat Island, Abu Dhabi, UAE\\
$^{3}$Blue Marble Space Institute of Science, 1001 4th Ave., Suite 3201, Seattle, WA, 98154, USA
}

\date{Accepted XXX. Received YYY; in original form ZZZ}

\pubyear{2019}

\begin{document}
\label{firstpage}
\pagerange{\pageref{firstpage}--\pageref{lastpage}}
\maketitle

\begin{abstract}
The discovery of terrestrial exoplanets orbiting in habitable zones around nearby stars has been one of the significant developments in modern astronomy. More than a dozen such planets, like Proxima Centauri b and TRAPPIST-1 e, are in close-in configurations and their proximity to the host star makes them highly sensitive to stellar activity. Episodic events such as flares have the potential to cause severe damage to close-in planets, adversely impacting their habitability. Flares on fast rotating young M stars occur up to 100 times more frequently than on G-type stars which makes their planets even more susceptible to stellar activity. Stellar Energetic Particles (SEPs) emanating from Stellar Proton Events (SPEs) cause atmospheric damage (erosion and photochemical changes), and produce secondary particles, which in turn results in enhanced radiation dosage on planetary surfaces. We explore the role of SPEs and planetary factors in determining planetary surface radiation doses. These factors include SPE fluence and spectra, and planetary column density and magnetic field strength. Taking particle spectra from 70 major solar events (observed between 1956 and 2012) as proxy, we use the GEANT4 Monte Carlo model to simulate SPE interactions with exoplanetary atmospheres, and we compute surface radiation dose. We demonstrate that in addition to fluence, SPE spectrum is also a crucial factor in determining the surface radiation dose. We discuss the implications of these findings in constraining the habitability of terrestrial exoplanets. 
\end{abstract}


\section{Introduction}
Over the past two decades we have seen a surge in the discovery of exoplanets. At present about 4000 exoplanets have been detected and the number is growing at an ever-increasing pace \footnote{NASA Exoplanet Archive (Caltech/NASA)}. As we learn more about exoplanets and their host stars, one of the most important question is: are any of the known exoplanets habitable? The Circumstellar Habitable Zone (CHZ) \citep{kasting1993habitable} is the area around a star where, based on reasonable assumptions about the planetary atmosphere, the stellar flux is sufficient to provide a temperature suitable for maintaining liquid water on its surface. A number of exoplanets, such as such as Proxima Centauri b and TRAPPIST-1 e have been found orbiting in CHZ in close proximity to the host star \footnote{Habitable Exoplanets Catalog (2019) \url{phl.upr.edu/hec}}. This proximity, however, also makes them highly sensitive to stellar activity and might adversely affect their habitability. 

Space weather events, such as stellar flares, CMEs, and Stellar Proton Events (SPEs), abruptly bombard the planet with non-thermal radiation consisting of X-rays, EUV (XUV), and Stellar Energetic Particles (SEPs). Space weather has a significant impact on planets, with atmospheric effects such as erosion \citep{jakosky2015maven}, photochemical changes \citep{atri2010lookup, segura2010effect, tabataba2016atmospheric, scheucher2018new}, and enhanced radiation dose on the surface \citep{atri2014cosmic, atri2016modelling, guo2018generalized}. Earlier studies have shown that stellar flares have the capacity to significantly alter habitable conditions on planets \citep{airapetian2017hospitable, atri2016modelling, lingam2017risks}. Flares with energies up to 10$^{36}$ ergs (superflares) have been observed from {\it Kepler} and {\it Gaia} observations, and estimates of their energy and frequency on different types of stars is improving rapidly \citep{schaefer2000superflares, maehara2012superflares, davenport2016kepler, 2019ApJ...876...58N}. Flares with energies up to 10$^{35}$ ergs occur about once every 2000-3000 years on slow rotating stars like the Sun, but the occurrence rate is $\sim$ 100 times higher for younger, faster rotating stars of the same class \citep{2019ApJ...876...58N}. 

Charged particle emission from Solar space weather events and their impact on the Earth have been studied in great detail; these studies serve as a great resource to estimate similar effects on exoplanets. The trajectory of charged particles is determined by the planetary magnetosphere which can thus shield the planet against SEP-induced effects \citep{griessmeier2015galactic}. Charged particles interact with the planetary atmosphere, undergo hadronic interactions \citep{beringer2012review}, leading to atmospheric changes (ionization, photochemistry, erosion) \citep{melott2011astrophysical, atri2014cosmic, tabataba2016atmospheric} and enhanced surface radiation dose \citep{atri2016modelling}, which could impact the habitats of potential ecosystems on the planet. Therefore, the planetary radiation environment in the higher energy regime (MeV - GeV range) is an important factor determining the habitability of an exoplanet. 

While most research on the subject has been focused on atmospheric effects of flares \citep{segura2010effect} and its indirect effect on life, this {\it Letter} is focused on estimating direct damage to life from flare-induced radiation. We extend our earlier work on the impact of SPEs on terrestrial exoplanets \citep{atri2016modelling} by incorporating data from 70 major SPEs and focus on the impact of spectral shape on the surface radiation dose. We calibrate our model with surface radiation dose observations on Mars \citep{lee2018observations, guo2018generalized} and compute SEP-induced radiation dose enhancement on exoplanetary surfaces. 

\section{Method}
Although observational data of stellar flare photon emission is widely available, at present there are no observations of charged particle emission from stars other than the Sun. In absence of this data, we rely on major solar SPEs (observed between 1956 and 2012) \citep{tylka2009new}, which have been studied in great detail, as proxies of stellar events. NOAA's GOES mission \citep{sandberg2014cross} has been instrumental in measuring charged particle emission from solar events and has been a reliable source of data for decades. The energy of protons emitted from these events ranges roughly between 10 MeV and 10 GeV \citep{tylka2009new}. Although, lower and higher energy particles have been observed, this energy range is widely used for SPEs. The duration of events varies from a few hours to a few days in case of most extreme events. Energetic protons are capable of producing secondary particles in the atmosphere, sometimes resulting in GLEs (Ground Level Enhancements). GLEs are detected with a wide range of instruments such as neutron monitors, muon detectors, and other ground-based instruments spread throughout the globe. Data from all major solar events has been carefully analyzed to obtain the time-averaged spectrum of each event and is available in parametrized form \citep{tylka2009new}. Since time evolution of the SPE spectrum is not relevant to this {\it Letter} as in case of photochemistry or climate-related calculations, the event-integrated averaged spectrum of each event was used for calculations (we focus on total radiation dose per event).  

Charged particles with energies above the pion production threshold (290 MeV) \citep{beringer2012review} undergo hadronic interactions upon interaction with the atmosphere in addition to electromagnetic interactions; simple analytical solutions are insufficient to model the interactions \citep{beringer2012review}. GEANT4 is a Monte Carlo package developed at CERN to model energetic charged particle interactions. It has been extensively calibrated with a variety of experiments worldwide \citep{agostinelli2003geant4}. The tool is also used in planetary sciences to model charged particle interactions with the atmosphere \citep{atri2016modelling, guo2018generalized}. We use GEANT4 to model the interaction of SEPs in the atmosphere, and we compute the radiation dose deposited at the ground level for different atmospheric depths (column density). Each SPE was simulated with 10$^{9}$ protons in the energy range of 10 MeV to 10 GeV given by the parametrized spectra \citep{tylka2009new}. The typical fluence of the 70 events considered here is also $\sim$ 10$^{9}$ protons cm$^{-2}$ \citep{tylka2009new}. The upper limit of recorded fluence is $\sim$ 10$^{10}$ protons cm$^{-2}$ for the Carrington event \citep{smart2006carrington}. The code tracks one particle at a time, and as a result, for the same spectrum, the radiation dose scales linearly with fluence \citep{agostinelli2003geant4}. The radiation dose is primarily dependent on the overall column density of the atmosphere whereas the atmospheric composition is not crucial as in case of photochemistry or atmospheric loss \citep{atri2013galactic,atri2016did}. Since little is known about the atmospheres of terrestrial exoplanets, parameters from the Earth's atmosphere were used (along with mass, radius, and gravity), except for the atmospheric depth (1036 g cm$^{-2}$), which is a free parameter in this study. One important factor determining the flux of charged particles entering the planet's atmosphere is its magnetosphere. We used magnetospheric modeling results relevant to this problem from \cite{griessmeier2015galactic} which provided us with filter functions for different magnetospheric strengths. These filter functions determine the probability that a charged particle of a particular energy will enter the atmosphere. Radiation dose was calculated using the method described in \cite{atri2013galactic, atri2016modelling}. 

We have validated our model by computing the GCR-induced background radiation dose on Mars measured by the Radiation Assessment Detector (RAD) on board the Mars Science Laboratory (MSL). We used the BON10 model \citep{o2010badhwar} to obtain the background GCR spectrum, with 87\% protons, 12\% alpha particles and 1\% Iron, as a substitute for heavier particles and used the MCD model \citep{forget1999improved, millour2015mars} for the Martian atmosphere. The background rate measured by RAD was found to be 210$\pm$40 $\mu$Gy/day \citep{hassler2014mars}; our model furnished a background rate of 218.5 $\mu$Gy/day, consistent with the RAD data within the limits of uncertainty in measurements. 

Since stellar flares can be several orders of magnitude more energetic than observed solar flares \citep{maehara2012superflares, davenport2016kepler, 2019ApJ...876...58N}, it is important to know the relation between flare energy (photons) and total particle fluence for events which go well beyond observations. Due to lack of data we will use a simple linear scaling relation we used earlier in \cite{atri2016modelling} to scale the proton fluence to extreme events. We then compare it with other methods based on theoretical models. As will be demonstrated in the following, the radiation dose on a planetary surface mainly depends on four factors: stellar particle fluence, particle spectrum, planetary atmospheric depth, and the magnetic field strength. In order to decouple the effect of fluence and spectrum, the fluence is normalized for all 70 events at 10$^{9}$ protons cm$^{-2}$ , with no changes in the event spectral shape, so that the distribution of surface radiation dose as a function of spectra can be obtained. Radiation dose is given in units of Gray (1 Gy = 1 J/kg), and the magnetic field strength is relative to the Earth's magnetic field strength.

\section{Results}
 As shown in Figure 1, for a constant fluence of 10$^{9}$ protons cm$^{-2}$, the radiation dose varies by about 5 orders of magnitude as a function of the event number. This is due to the fact that lower energy particles (MeV range) do not contribute to surface radiation dose, and the dose depends only on the fluence of particles in the higher end of the spectrum ($\geq$ 0.5 GeV). The results indicate that planets at a semi-major axis of 1 AU are not exposed to high levels of radiation, $\geq$1 Gy (considered harmful for organisms such as mammals \citep{real2004effects}), for typical flares even with minimal atmospheres and magnetic field strengths (Tables 1 and 2). Tables 1 and 2 can be used to compute radiation dose for any system by scaling the values for the orbital distance of the planet and SPE fluence. Table 3 shows results for a large-fluence (10$^{11}$ protons cm$^{-2}$) median-spectrum event on Proxima b and TRAPPIST-1 system planets and it can be seen that such an event does not pose any significant threat to potential ecosystems. Tables 4 and 5 show results from a large-fluence (10$^{11}$ protons cm$^{-2}$) hard-spectrum event on potentially habitable exoplanets with Earth's magnetic field strength (Table 4) and without any magnetic field (Table 5). The hard-spectrum event is the 24 August 1998 event, which had low fluence but gives the highest dose after normalization. Such an event could blast the planet with levels of radiation dose (1-10 Gy) that could be lethal to life forms, including mammals. The dose estimates at a distance of 1 km from the hypocenter at ground ground level in Hiroshima and Nagasaki were 7 Gy and 10 Gy respectively \citep{cullings2006dose}. This could cause mass extinction for a terrestrial-type biosphere. For a hard-spectrum event, the spectral shape is dominated by particles of several GeVs, 10 GeV being the upper limit, and such particles are modulated by a lesser extent by the planetary magnetic field, compared to lower energy particles. This effect can also be seen in Tables 4 and 5 (with and without the magnetic field). 

It should be noted that the surface radiation dose can be several orders of magnitude higher for superflares. Superflares are significantly more frequent on fast rotating young stars than they are on slow-rotating, solar-age stars \citep{2019ApJ...876...58N}, which will result in a higher cumulative radiation dose. As discussed earlier, flares with energy $\sim$ 10$^{35}$ erg occur about once every 2000-3000 years on slow rotating stars like the Sun, but the occurrence rate is $\sim$ 100 times higher for younger, faster rotating stars of the same class \citep{2019ApJ...876...58N}. This would make life on such planets vulnerable to frequent extinction-level radiation blasts by terrestrial standards with only highly radioresistant extremophiles surviving ($\sim$ 100 Gy). Such levels are expected for close-in planets such as Proxima b (Figure 2) in extreme cases. We also obtain the background rate of GCR-induced radiation dose for each planet using the method described by \cite{atri2013galactic} and \cite{griessmeier2016galactic}, we find enhancement factor of radiation dose over GCR background (Tables 6 and 7). Table 6 gives the enhancement factor over GCR background and Earth's magnetic field, and Table 7 without a magnetic field. It can be seen that in some cases the enhancement of radiation dose can be over 4 orders of magnitude compared to the background rate from GCRs.  Finally, in Figure 2, we show an extreme case, the dose distribution on Proxima b with 30 g cm$^{-2}$ atmosphere and no magnetic field. The fluence is extrapolated linearly and as one can notice, the dose levels can be in 1000s of Gy in extreme cases. 

The TRAPPIST-1 system shows very high enhancements, which can be seen in Tables 3-7, and is of great interest to astrobiologists. \cite{Fraschetti_2019} calculated the rate of 10 GeV protons to be 1.2$\times$10$^{5}$ cm$^{-2}$ sr$^{-1}$ s$^{-1}$ at TRAPPIST-1e (0.028 AU), which equates to $\sim$ 10$^{9}$-10$^{10}$ protons cm$^{-2}$. For context, the fluence of a typical solar event is $\sim$ 10$^{9}$ protons cm$^{-2}$ \citep{tylka2009new} with a recorded upper limit of $\sim$ 10$^{10}$ protons cm$^{-2}$ for the Carrington event \citep{smart2006carrington}. We carried out simulations with 10 GeV monoenergetic protons for a 1 hr event and found the surface radiation dose for a planet with 30 g cm$^{-2}$ atmospheric depth to be 1.23 Gy and with 1000 g cm$^{-2}$ atmospheric depth to be 1.3$\times$10$^{-3}$ Gy. On the other hand, \cite{struminsky2018radiation} used the older Parker 1958 model, instead of the widely used Parker 1965 model, and estimated a maximum fluence of 10$^{12}$-10$^{13}$ protons cm$^{-2}$ in 30-200 MeV energy range. Due to the lower energy range of particles, the event will have negligible impact on the surface radiation dose.
\begin{figure}
	\includegraphics[width=\columnwidth]{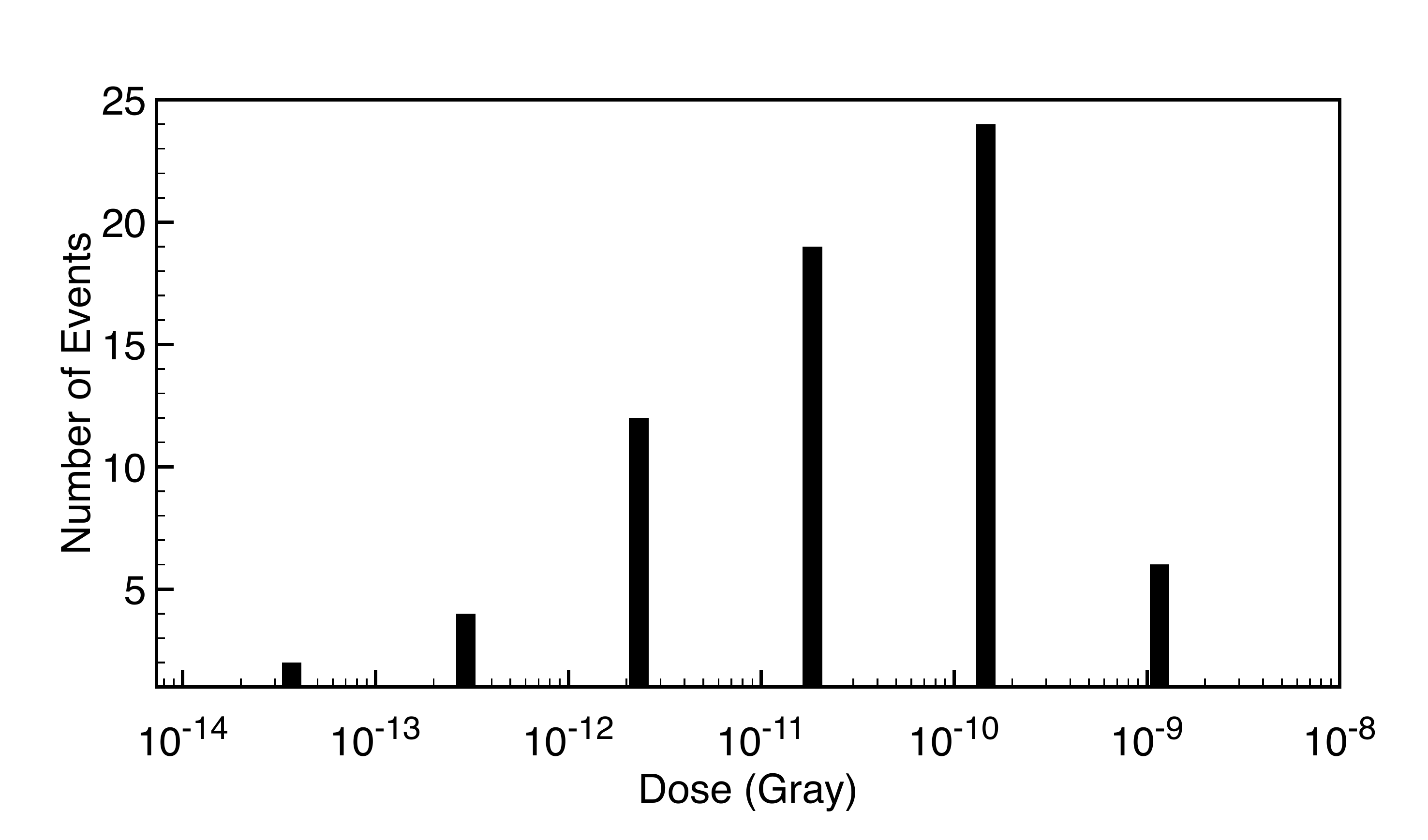}
    \caption{Radiation dose (Gy) distribution at 1 AU with 1000 g cm$^{-2}$ atmosphere (Earth-like) and Earth's magnetic field strength.}
    \label{fig:Dose distribution}
\end{figure}

\begin{table}
	\centering
	\caption{Median radiation dose (Gy) for various scenarios with a normalized fluence of 10$^{9}$ protons cm$^{-2}$ for a planet at 1 AU. Atmospheric depth varies between 30 and 1000 g cm$^{-2}$ (Earth's atmospheric depth is 1036 g cm$^{-2}$) and magnetospheric strength from 0.15 to 10 times the Earth value.}
	\label{tab:example_table1}
	\begin{tabular}{lccccr} 
		
		\hline
		  & 0.15 & 0.5 & 1 & 3 & 10\\
		\hline
		30 & 1.54E-08 & 8.06E-09 & 4.78E-09 & 1.94E-09 & 1.74E-10\\
		70 & 1.20E-08 & 6.42E-09 & 3.80E-09 & 1.60E-09 & 1.50E-10\\
		100 & 9.63E-09 & 5.24E-09 & 3.09E-09  & 1.34E-09 & 1.32E-10\\
		300 & 2.22E-09 & 1.33E-09 & 7.80E-10 & 3.45E-10 & 4.18E-11\\
		700 & 1.51E-10 & 8.70E-11 & 5.13E-11 & 2.17E-11 & 2.79E-12\\
		1000 & 8.15E-12 & 4.18E-12 &	2.49E-12 & 9.73E-13 & 1.01E-13\\
		\hline
	\end{tabular}
\end{table}

\begin{table}
	\centering
	\caption{Maximum radiation dose (Gy) for various scenarios with a normalized fluence of 10$^{9}$ protons cm$^{-2}$ for a planet at 1 AU. Atmospheric depth varies between 30 and 1000 g cm$^{-2}$ (Earth's atmospheric depth is 1036 g cm$^{-2}$) and magnetospheric strength from 0.15 to 10 times the Earth value.}
	\label{tab:example_table2}
	\begin{tabular}{lccccr} 
		
		\hline
		  & 0.15 & 0.5 & 1 & 3 & 10\\
		\hline
		30 & 1.76E-04	&9.99E-05&	5.89E-05&	2.54E-05&	3.02E-06\\
		70 & 1.42E-04&	8.18E-05	&4.81E-05	&2.11E-05&	2.61E-06\\
		100 & 1.17E-04	&6.85E-05	&4.02E-05	&1.79E-05	&2.29E-06\\
		300 & 2.91E-05	&1.82E-05&	1.07E-05	&4.90E-06&	7.22E-07\\
		700 & 1.89E-06	&1.18E-06&	6.95E-07&	3.08E-07&	4.87E-08\\
		1000 & 9.16E-08&	5.24E-08	&3.10E-08&	1.30E-08&	1.76E-09\\
		\hline
	\end{tabular}
\end{table}
\begin{table}
	\centering
	\caption{Radiation dose (Gy) for 10$^{11}$ protons cm$^{-2}$ event for Proxima b and TRAPPIST-1 system planets. Atmospheric depth varies between 30 and 1000 g cm$^{-2}$.}
	\label{tab:example_table3}
	\begin{tabular}{lcccr} 
		
		\hline
		  		  & 30 &  100 & 300 &  1000 \\

		\hline
		Proxima b & 6.55E-04&		4.09E-04&	9.43E-05&	3.46E-07\\
		TRAPPIST-1 b & 1.25E-02 & 7.80E-03 & 1.80E-03 & 6.60E-06 \\
		c & 6.66E-03 & 4.16E-03 & 9.59E-04  & 3.52E-06 \\
		d & 3.35E-03 & 2.09E-03 & 4.83E-04 & 1.77E-06 \\
		e & 1.94E-03&	1.21E-03	&2.80E-04	&	1.03E-06 \\
		f & 1.12E-03&		6.99E-04	&1.61E-04	&	5.92E-07\\
		g&7.57E-04&	4.73E-04&	1.09E-04	&	4.01E-07\\
		h&3.88E-04	&	2.43E-04	&5.59E-05	&	2.05E-07\\
		\hline
	\end{tabular}
\end{table}
\begin{table}
	\centering
	\caption{Radiation dose (Gy) on potentially habitable planets for a hard spectrum event (24 August 1998) with 10$^{11}$ protons cm$^{-2}$ fluence and Earth's magnetic field. Atmospheric depth varies between 30 and 1000 g cm$^{-2}$.}
	\label{tab:example_table4}
	\begin{tabular}{lccccr} 
		
		\hline
		  & d (AU) & 30  &  100  & 300 &  1000 \\
		\hline
		TRAPPIST-1 e & 0.028 & 7.42E+00&5.07E+00&	1.35E+00&	3.90E-03\\
		TRAPPIST-1 f & 0.037 & 4.28E+00&2.92E+00&	7.76E-01&		2.25E-03\\
		TRAPPIST-1 g&0.045&	2.89E+00&		1.98E+00&	5.25E-01&	1.52E-03\\
		Proxima Cen b & 0.049 & 2.50E+00&	1.71E+00&	4.54E-01&	1.32E-03 \\
		GJ 667 C f & 0.156 & 2.42E-01&	1.65E-01&	4.39E-02&	1.27E-04 \\
		GJ 667 C e & 0.213 & 1.30E-01&	8.87E-02&	2.35E-02&	6.83E-05\\
        Kepler-1229 b & 0.301 & 6.52E-02	&	4.45E-02&	1.18E-02&	3.43E-05 \\
        Kepler-442 b & 0.409 & 3.52E-02&	2.41E-02&	6.38E-03&	1.85E-05 \\
        Kepler-186 f&0.432&	3.15E-02		&2.16E-02&	5.72E-03&1.66E-05\\
        Kepler-62 f& 0.718 & 1.14E-02&	7.81E-03&	2.07E-03&		6.01E-06\\
		
		\hline
	\end{tabular}
\end{table}
\begin{table}
	\centering
	\caption{Radiation dose (Gy) on potentially habitable planets for a hard spectrum event (24 August 1998) with 10$^{11}$ protons cm$^{-2}$ fluence and no magnetic field. Atmospheric depth varies between 30 and 1000 g cm$^{-2}$.}
	\label{tab:example_table5}
	\begin{tabular}{lccccr} 
		
		\hline
		  & d (AU) & 30  &  100  & 300 &  1000 \\
		\hline
TRAPPIST-1 e& 0.028&	2.22E+01&	1.47E+01	&3.67E+00&		1.15E-02\\
TRAPPIST-1 f& 0.037&	1.28E+01	&	8.47E+00	&2.12E+00	&	6.66E-03\\
TRAPPIST-1 g&0.045&	8.67E+00	&	5.73E+00&	1.43E+00	&	4.50E-03\\
Proxima Cen b&0.049&	7.50E+00	&	4.96E+00	&1.24E+00	&	3.89E-03\\
GJ 667 C f	&0.156& 7.25E-01	&	4.79E-01	&1.20E-01&	3.76E-04\\
GJ 667 C e	&0.213& 3.89E-01	&	2.57E-01&	6.42E-02	&	2.02E-04\\
Kepler-1229 b&0.301& 	1.95E-01	&	1.29E-01&	3.22E-02	&	1.01E-04\\
Kepler-442 b&0.409&	1.05E-01	&	6.97E-02&	1.74E-02	&	5.48E-05\\
Kepler-186 f&0.432& 9.45E-02	&	6.25E-02&	1.56E-02	&	4.91E-05\\
Kepler-62 f	&0.718& 3.42E-02	&	2.26E-02&	5.65E-03	&	1.78E-05\\

		\hline
	\end{tabular}
\end{table}
\begin{table}
	\centering
	\caption{Enhancement factor over GCR background on potentially habitable planets for a hard spectrum event (24 August 1998) with 10$^{11}$ protons cm$^{-2}$ fluence and Earth's magnetic field. Atmospheric depth varies between 30 and 1000 g cm$^{-2}$. }
	\label{tab:example_table6}
	\begin{tabular}{lcccr} 
		
		\hline
		  & 30  &  100  & 300 &  1000 \\
		\hline
		TRAPPIST-1 e & 1.35E+04&	1.08E+04&	1.03E+04&	2.19E+03\\
		TRAPPIST-1 f&7.81E+03	&6.23E+03	&5.97E+03	&1.26E+03\\
		TRAPPIST-1 g&5.28E+03	&4.22E+03	&4.04E+03	&8.56E+02\\
		Proxima Cen b & 4.57E+03&	3.65E+03&	3.49E+03&	7.40E+02 \\
		GJ 667 C f b & 4.41E+02	&3.53E+02	&3.37E+02	&7.15E+01 \\
		GJ 667 C e & 2.37E+02	&1.89E+02	&1.81E+02	&3.84E+01\\
		Kepler-1229 b & 1.19E+02&	9.50E+01&	9.09E+01&	1.93E+01 \\
		Kepler-442 b & 6.42E+01	&5.13E+01	&4.91E+01	&1.04E+01 \\
		Kepler-186 f	&5.76E+01	&4.60E+01	&4.40E+01	&9.33E+00\\
		Kepler-62 f&2.08E+01	&1.66E+01	&1.59E+01	&3.38E+00\\

		\hline
	\end{tabular}
\end{table}
\begin{table}
	\centering
	\caption{Enhancement factor over GCR background on potentially habitable planets for a hard spectrum event (24 August 1998) with 10$^{11}$ protons cm$^{-2}$ fluence and no magnetic field. Atmospheric depth varies between 30 and 1000 g cm$^{-2}$. }
	\label{tab:example_table7}
	\begin{tabular}{lcccr} 
		
		\hline
		  & 30  &  100  & 300 &  1000 \\
		\hline
TRAPPIST-1 e&1.16E+04	&9.73E+03	&9.51E+03	&6.49E+03\\
TRAPPIST-1 f&	6.67E+03	&5.61E+03	&5.48E+03	&3.74E+03\\
TRAPPIST-1 g&	4.52E+03	&3.80E+03	&3.71E+03	&2.53E+03\\
Proxima Cen b&	3.90E+03	&3.28E+03	&3.21E+03	&2.19E+03\\
GJ 667 C f	&3.77E+02&	3.17E+02	&3.10E+02	&2.11E+02\\
GJ 667 C e	&2.02E+02&	1.70E+02	&1.66E+02	&1.13E+02\\
Kepler-1229 b&	1.02E+02	&8.55E+01	&8.35E+01	&5.70E+01\\
Kepler-442 b&	5.49E+01	&4.62E+01	&4.51E+01	&3.08E+01\\
Kepler-186 f	&4.92E+01	&4.14E+01	&4.04E+01	&2.76E+01\\
Kepler-62 f	&1.78E+01&	1.50E+01	&1.46E+01	&9.98E+00\\

		\hline
	\end{tabular}
\end{table}

\begin{figure}
	\includegraphics[width=\columnwidth]{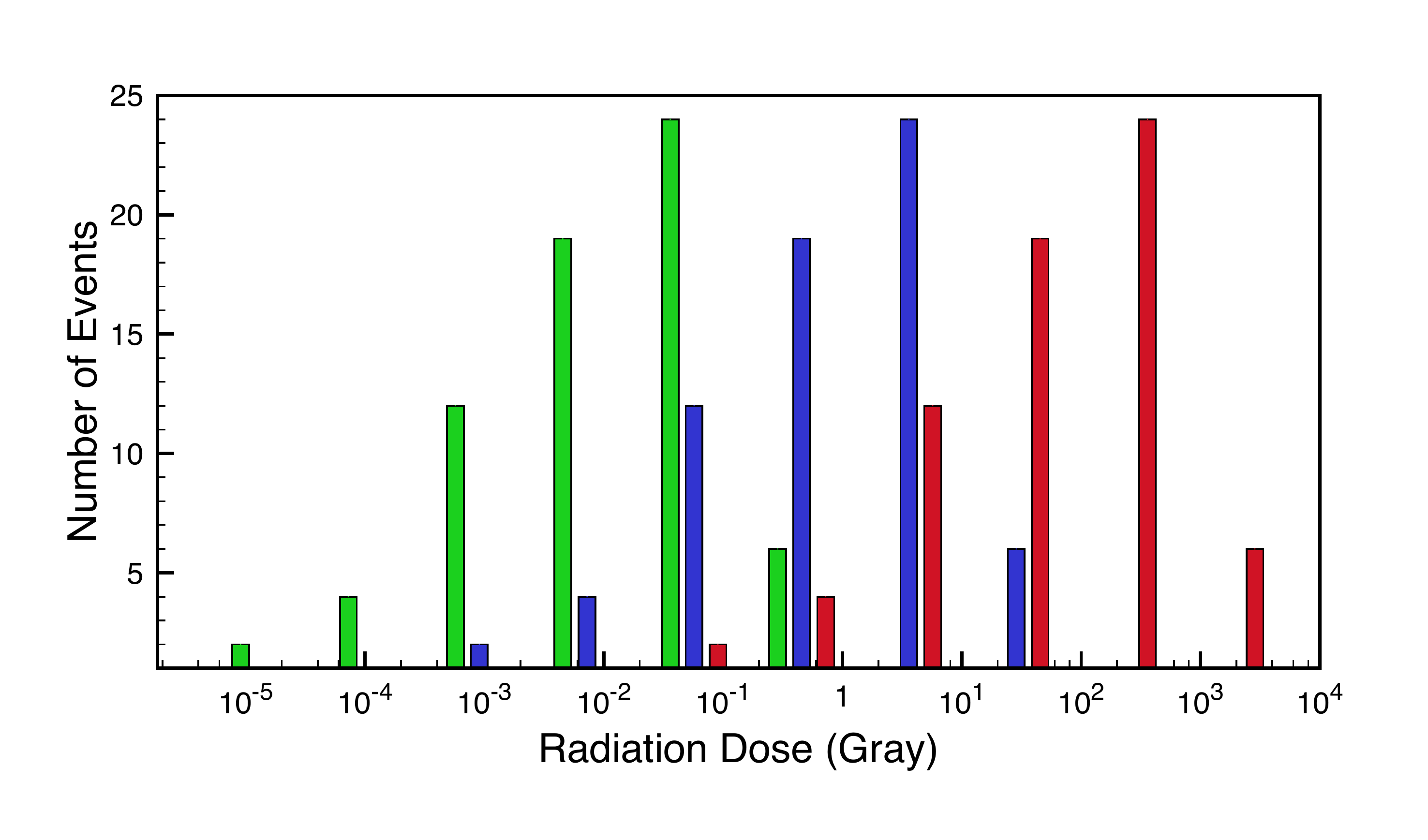}
    \caption{Dose distribution on Proxima b with 30 g cm$^{-2}$ atmosphere and no magnetic field corresponding to 10$^{11}$ (green), 10$^{13*}$ (blue) and 10$^{15*}$ (red) protons cm$^{-2}$ fluence (* Linear extrapolation, not observation based).}
    \label{fig:proxima b}
\end{figure}

\section{Impact on Habitability}
Space weather can have a profound impact on planetary surfaces and as we have shown, has the potential to disrupt hospitable conditions on terrestrial exoplanets. We found that SPEs on close-in terrestrial planets can significantly enhance the surface radiation dose and adversely impact their habitability. Flare frequency, spectrum, magnetic field, and particle fluence govern the planetary radiation environment in the higher energy regime considered in this work. The occurrence rate of superflares on fast rotating young M stars (rotation period $\sim$ a few days, age $\sim$ a few 100 Myr) is about 100 times higher than that on G-type stars \citep{2019ApJ...876...58N}. Sun-like stars can have superflares up to 5$\times$10$^{34}$ ergs every 2000-3000 years \citep{2019ApJ...876...58N}. 

The radiation dose on the planetary surface is determined by its atmospheric column density and to a lesser extent its magnetic field. Direct exposure to ionizing radiation can have a harmful influence \citep{ferrari2009cosmic, melott2011astrophysical} on potential ecosystems of habitable exoplanets. Abrupt bursts of radiation can damage or incapacitate organisms and significantly alter their habitats \citep{atri2014cosmic, lingam2017risks}. In planetary systems with a high occurrence rate of flares, radiation from SPEs can be a crucial factor determining the type of life, if any that could exist on a planet \citep{atri2016modelling}. In addition to strong exposure to SEPs, planets with minimal atmospheres and/or magnetic field strength experience heightened background dose from Galactic Cosmic Rays \citep{atri2013galactic, griessmeier2015galactic, griessmeier2016galactic}. 

These estimates can be improved with better theoretical models estimating SEP fluence and frequency \citep{struminsky2018radiation, Fraschetti_2019}. We have used a linear scaling relation of SPE particle fluence for our calculations but its validity for superflares needs to be studied further along with particle ejection models for active stars, some of which suggest that magnetically active stars might hinder particle emission in superflares \citep{alvarado2018suppression, Moschou_2019}. More progress in this area will help us better understand the relation between radiation dose from flares and its impact on planetary habitability. Irrespective of the type of life which could potentially exist on other planets (carbon-based or not), such life would be exposed to significantly enhanced dose rate as calculated in this {\it Letter}. Radioresistance might be the most important characteristic of surviving organisms which will presumably have the ability to effectively repair damage caused by ionizing radiation \citep{minton1994dna}. In such an environment, organisms need an efficient repair mechanism (eg. radioresistant bacterium {\it Deinococcus radiodurans}), since such an environment requires use of a significant amount of energy in repairing damage, and leaving less energy for basic metabolic activities, growth, and reproduction \citep{makarova2001genome}. It is therefore likely that extremophiles surviving in high radiation environment will have slow metabolisms and growth rates (The most radioresistant extremophile {\it D. radiodurans} grows only after the radiation exposure stops \citep{makarova2001genome}), and radiation can act as a constraint on the type of organisms which could exist in such environments. Subsurface habitats will be shielded from this radiation depending on the column density of rock or water above it \citep{atri2016possibility}. On the other hand, moderate enhancements in radiation levels can increase mutation rates and may lead to accelerated evolution and richer biodiversity. If the mutation rates are high, it can act as a limiting factor in evolutionary adaptation \citep{sprouffske2018high}. It can also be argued that on initially lifeless planets, increased radiation dose can assist in the production of organic substances including prebiotic molecules and may play a role in the origin of life \citep{atri2016possibility, lingam2018propitious}. 

\section{Conclusions}
SPEs can abruptly enhance radiation dose on planetary surfaces and have the capability to disrupt potentially habitable conditions on planets. We have demonstrated that radiation dose varies significantly with charged particle spectra and an event of a given fluence can have a drastically different effect depending on the spectrum. Our results show that radiation dose can vary by about 5 orders of magnitude for a given fluence. In terms of shielding, we found that atmospheric depth (column density) is a major factor in determining radiation dose on the planetary surface. Radiation dose is reduced by 3 orders of magnitude corresponding to an increase in the atmospheric depth by an order of magnitude. We found that the planetary magnetic field is an important but a less significant factor compared to atmospheric depth. The dose is reduced by a factor of about 30 corresponding to an increase in the magnetospheric strength by an order of magnitude. However, it should be noted that planetary magnetic field is crucial in maintaining a substantial atmosphere on a planet \citep{jakosky2015maven, griessmeier2004effect, khodachenko2007mass, lammer2007coronal, driscoll2013divergent, dong2017seasonal, gunell2018intrinsic}. Although recent observations have given us good measurements of flaring rates of nearby stars \citep{maehara2012superflares, davenport2016kepler, 2019ApJ...876...58N}, the main source uncertainty in this work is the lack in measurement of particles ejected by high-energy flares (10$^{32}$ - 10$^{36}$ ergs) on other stars. More progress in this area will improve our understanding of the relationship between extreme solar events, radiation dose and planetary habitability.

\section*{Acknowledgements}

This work is supported by the NYUAD Institute research grant G1502. I thank the anonymous reviewer for their detailed feedback which improved the manuscript, members of the GEANT4 collaboration who developed the software used for this work, and Philip Rodenbough of the NYUAD Scientific Writing Program for his feedback. This research was carried out on the HPC resources at NYUAD. 

\bibliographystyle{mnras}
\bibliography{spe_2019} 

\label{lastpage}
\end{document}